# ULTRACOOL DWARFS IN GAIA

## C. REYLÉ

Université de Franche-Comté, Institut UTINAM, CNRS UMR6213, OSU THETA Franche-Comté-Bourgogne, Observatoire de Besançon, BP 1615, 25010 Besançon Cedex, France
e-mail: celine.reyle@obs-besancon.fr

**Abstract.** The Gaia optical observations are not the most suitable spectral domain for studying the low-mass, faintest and reddest part of the main sequence. Nevertheless, the large number of objects observed with an unprecedented precision, including trigonometric parallax, makes it an unrivaled dataset for studying the stellar-substellar boundary. In this paper, I review the contribution of the successive catalogues offered by the Gaia mission to study the low-mass, ultra-cool objetcs. I also present further characterisations and scientific exploitations of the Gaia sample.

## 1 Introduction

Ultra-cool dwarfs (hereafter UCDs) are defined as M7 and later type objects by Kirkpatrick et al. (1997). Their temperature is below about 2700 K (e.g. Rajpurohit et al. 2013). They are the link between stars and brown dwarfs as they can be either the least massive stars or brown dwarfs, spanning the stellar-substellar masses. The UCD census is still incomplete even within 25 pc of the Sun: Bardalez Gagliuffi et al. (2019) estimated that 62% of M7-L5 are catalogued in this volume.

In what follows, I review the contribution of the *Gaia* space mission (Gaia Collaboration et al. 2016) on the study of UCDs. I first describe the UCD regime, in between the lowest mass stars and brown dwarfs, and give hints on the motivation to detect and characterize them. I illustrate the powerful utility of *Gaia*, thanks to its huge, homogeneous and precise dataset to reveal the stellar-substellar boundary. There is no doubt that *Gaia* next data releases will complete the picture.

## 2 Low-mass stars and browns dwarfs

A vast amount of energy is released via nuclear fusion occurring in the core of a star. The fusion reaction is accompanied by a net loss of mass, resulting in a release of energy. To fuse hydrogen, the core must have temperatures larger than 3 millions K. On the Main Sequence, the thermal pressure from fusion keeps a star from gravitational collapse and the star is in thermal and hydrostatic equilibrium. The stellar matter follows classical statistical physics: classical nearly perfect gas equation of state and quasi-static equilibrium condition. The radius of a star is proportional to its mass (see e.g. Figure 15 in Boyajian et al. 2012, and blue symbols in Figure 1). The less mass a star has, the more it needs to contract to heat the core, and the smaller it will be on the Main Sequence. M-dwarfs are the lowest mass stars, ten times smaller and ten times less massive than the Sun.

The question on how small can a star be and form by gravitational collapse of a giant molecular cloud is still open. The formation scenario, from filamentary clouds to pre-stellar cores to stars, has been refined in particular thanks to the wide-field *Herschel* photometric survey of nearby star-forming cloud complexes (André et al. 2010). However the gravitational collapse of a giant molecular cloud into very low mass objects requires a very high density (about $10^7$ cm$^{-3}$ to form a 0.07 M$_\odot$ object). Alternative mechanisms to form such small objects are still under debate: embryo ejection (Bate et al. 2002; Goodwin & Whitworth 2004; Reipurth & Mikkola 2015), photo-erosion of cores near massive stars (Whitworth & Zinnecker 2004), protostellar disc fragmentation (Vorobyov & Basu 2010; Attwood et al. 2009; Stamatellos & Herczeg 2015), gravoturbulent fragmentation (Padoan & Nordlund 2004; Hennebelle & Chabrier 2008; Bonnell et al. 2008; Lomax et al. 2016), or compression by turbulent flows in molecular clouds (Stamer & Inutsuka 2019).

The contraction of the forming object makes its core temperature and density increase. If its mass is not sufficient, the temperature will not reach the hydrogen burning threshold and the object will go towards the electron degeneracy limit. The collapse is stopped by electron degeneracy pressure: this substellar object is named a brown dwarf. This theoretical picture has been proposed in 1963 by Kumar (1963); Hayashi & Nakano (1963). Kumar (1963) estimated a hydrogen burning minimum mass of 0.07 $M_\odot$ (73 $M_{Jup}$) for objects with solar composition and 0.09 $M_\odot$ (94 $M_{Jup}$) for low metallicity objects. Brown dwarfs are the densest hydrogen-rich objects known (see Figure 1 from Hatzes & Rauer 2015). The macroscopic properties of the substellar matter are ruled by different physics and follow a different equation of state than in stars (e.g. Saumon et al. 1995; Chabrier et al. 2023).

Because brown dwarfs do not undergo stable hydrogen fusion, they cool down over time, progressively passing through later spectral types as they age. They are low-luminosity, very red objects, and difficult to detect. Strategies for finding these coolest objects are to search for them in young clusters (young brown dwarfs are brighter), to search for them as companions to other objects, to search for red objects in large scale surveys, eventually among large proper motion objects (being an indication of possible nearby, intrinsically faint, objects). First discoveries came out in the late 90s, thanks to the advance of the near-infrared technology: Teide 1 in the Pleiades (Rebolo et al. 1995), the cool brown dwarf Gliese 229B as a companion of the low-mass star Gliese 229A (Nakajima et al. 1995), Kelu 1 and three other free-floating brown dwarfs (Ruiz et al. 1997; Delfosse et al. 1997), and many discoveries since then thanks to the large scale surveys (DENIS, 2MASS, CFHTLS, SDSS, SIMP, UKIDSS, WISE, PanSTARRS,...).

In parallel to these numerous discoveries, evolutionary models were developed by several groups, assuming cloudless or cloudy atmospheres, placing the stellar-substellar limit between 70 and 79 $M_{Jup}$ (Chabrier & Baraffe 2000; Burrows et al. 2001, 2011; Saumon & Marley 2008; Baraffe et al. 2015; Marley et al. 2021). The empirical mass limit of hydrogen fusion has also been determined by Dupuy & Liu (2017) from the dynamical mass of 31 binaries. They estimated the stellar-substellar boundary at 70±4 $M_{Jup}$. In general, brown dwarf properties are bracketed between those of low-mass stars and massive giant planets. 13 $M_{Jup}$ is the mass limit to still allow nuclear fusion, that of deuterium (e.g. Burrows 1999). The IAU uses this mass as the limit to define a brown dwarf and an exoplanet, no matter how they formed (see Lecavelier des Etangs & Lissauer 2022).

The minimun size of a brown dwarf is about the size of Jupiter. In a more massive brown dwarf, gravitational force is higher and causes a larger fraction of the brown dwarf to become degenerate, causing it to have a smaller radius. The mass to radius relation shows a local minimum in the most massive brown dwarfs. The reversal of the mass-radius relation at the hydrogen burning limit is predicted by models (Figure 1). At a given mass, theoretical isochrones predict that older objects have smaller radii. Brown dwarfs and low mass stars observed by transit in binary systems, with age estimate from the primary star, are efficient to test the age-radius effect. Generally they validate model radii (Carmichael et al. 2021; Grieves et al. 2021).

Figure 2 (left panel) shows the evolution of objects with mass between 10 and 100 $M_{Jup}$ in the luminosity vs age diagram. The tracks are obtained from Baraffe et al. (2015) models. Once a star reaches the Main Sequence, its luminosity stabilizes, contrary to substellar objects. As a consequence, very low-mass stars, brown dwarfs, and planetary mass objects can have the same brightness (and effective temperature). As an illustration, a 2 Myr planetary mass object, a 50 Myr brown dwarf, or a 300 Myr star, have the same luminosity ($\sim 10^{-3}$ $L_\odot$). This is also clearly illustrated from observations of a sample of substellar companions with well-constrained ages and spectroscopically-derived classifications, in Figures 6 and 8 by Bowler (2016).

The evolution in effective temperature is also shown in Figure 2 (right panel). The horizontal lines at 2200 K, 1400 K, and 550 K roughly represent the transition between M and L dwarfs, L and T dwarfs, T and Y dwarfs, respectively. Spectral types of stars on the Main Sequence, O B F G K M, is a temperature (and also mass) related sequence. The spectral sequence for brown dwarfs, M L T Y, is not only a temperature sequence but also an evolutionary sequence. A massive, old, brown dwarf can have the same spectral type of a less massive, young, brown dwarf. The youngest brown dwarfs are M-types. Most of the brown dwarfs then evolve from M, L, and T-types. Only the less massive ones go to Y-types. The edge of the hydrogen burning main sequence is an L-dwarf.

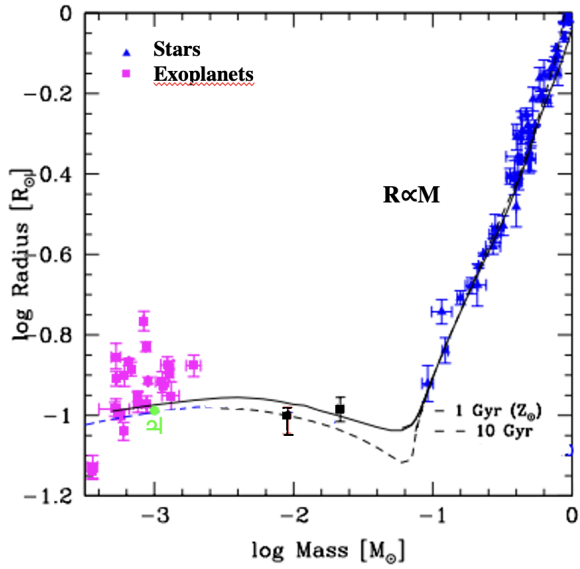

**Fig. 1.** Theoretical mass to radius relation computed from models at solar composition at 1 Gyr (solid line) and 10 Gyr (dashed line), superimposed to observationally-determined values (symbols). Jupiter is shown in green. Adapted from Chabrier et al. (2009).

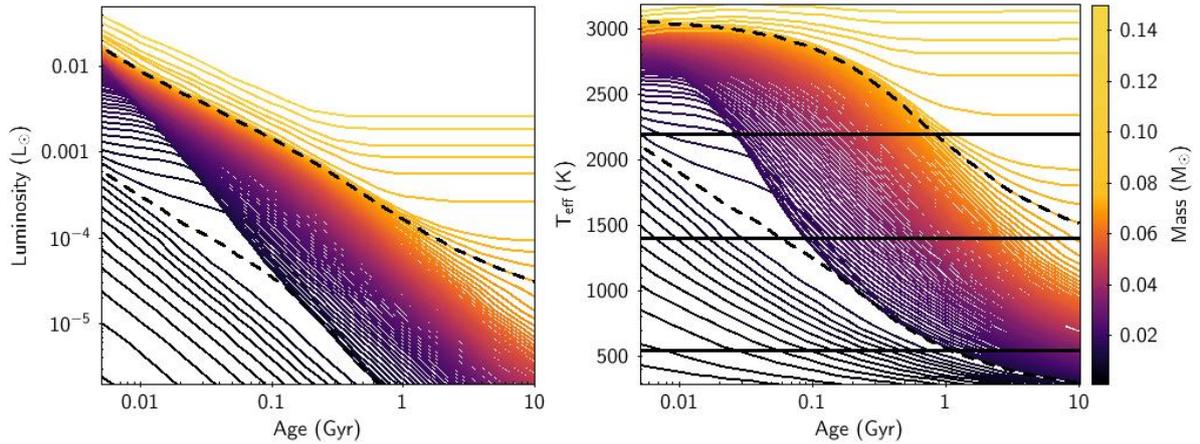

**Fig. 2.** Evolutionary tracks (left panel: luminosity, right panel: effective temperature) from Baraffe et al. (2015) models, for objects with planetary to stellar mass. The dotted lines are the 73 and 13 $M_{Jup}$ tracks. The horizontal lines (right panel) show the M/L, L/T, and T/Y transitions, from top to bottom.

## 3  Why are low-mass stars and brown dwarfs interesting to study?

Low-mass stars are the most numerous in the Galaxy. In the 10 pc volume-limited sample compiled by Reylé et al. (2021, 2022), 61% of the stars are M-dwarfs, half of them with a spectral type from M3 to M5. The *Gaia* Catalogue of Nearby Stars (GCNS, Gaia Collaboration et al. 2021b) is volume-limited to 100 pc, compiled from the *Gaia* early third data release (EDR3, Gaia Collaboration et al. 2021a). It contains at least 92% of stars of stellar type M9 within 100 pc of the Sun. The distribution of the stars peaks at the partly to fully convective stars boundary (Figure 3). The M-dwarfs ($8.25 \leq M_G \leq 15.5$) account for 64% of the full sample.

Low-mass stars and brown dwarfs evolve slowly and have a lifetime much larger than the age of the universe. Thus they span all ages and can be found in all populations, from young stellar associations to the halo (see section 5). They cover a huge range of astrophysical properties. Their absolute magnitudes span 20 magnitudes and 5 magnitudes in colors! This large variety of characteristics is the consequence of the complex physical processes acting in their cool atmospheres.

They remain elusive due to their faint magnitude. Modeling their complex, cool atmosphere is still a challenge. The strong absorptions due to the formation of molecules make their spectrum to deviate from the black body spectrum, and one cannot talk anymore of continuum but pseudo-continuum (Figure 4, left panel). In particular, the L-dwarfs offer a tricky and puzzling case, as illustrated in Figure 4, right panel. All classified as L3-dwarfs, the shape of their spectrum are modified by the condensates that form in their atmosphere. In the near infrared, dust can lead to back warming of the atmosphere, and alters the amount of $H_2O$ and $H_2$, and slight variations in the $H_2O$ and $H_2$ opacities can lead to large differences. Moreover, low-gravity objects will show a triangular shaped H-band (Kirkpatrick et al. 2008) whereas low-metallicity objects will present a higher flux in the J-band while suppressed H and K-bands. The full description of the atmosphere depends on effective temperature, gravity, metallicity, cloud properties, and mixing in the atmosphere.

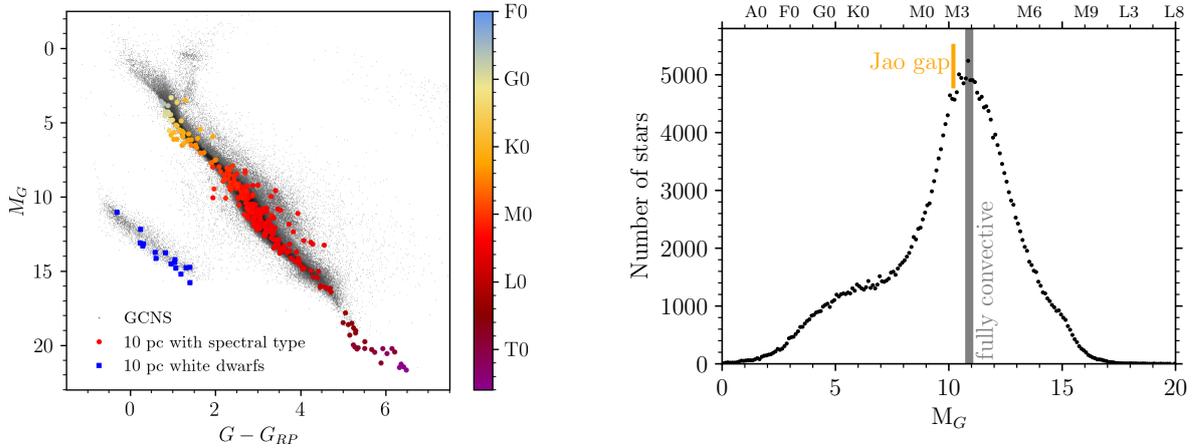

**Fig. 3.** Left panel: color absolute magnitude diagram of the GCNS (grey dots) superimposed with the 10 pc sample, colored by spectral type. Right panel: number of main sequence stars, in 0.1 magnitude bins, as a function of absolute magnitude, obtained from the GCNS. The high quality and precision of the sample allow to detect the Jao gap (Jao et al. 2018), a structure in the main sequence related to structural instabilities.

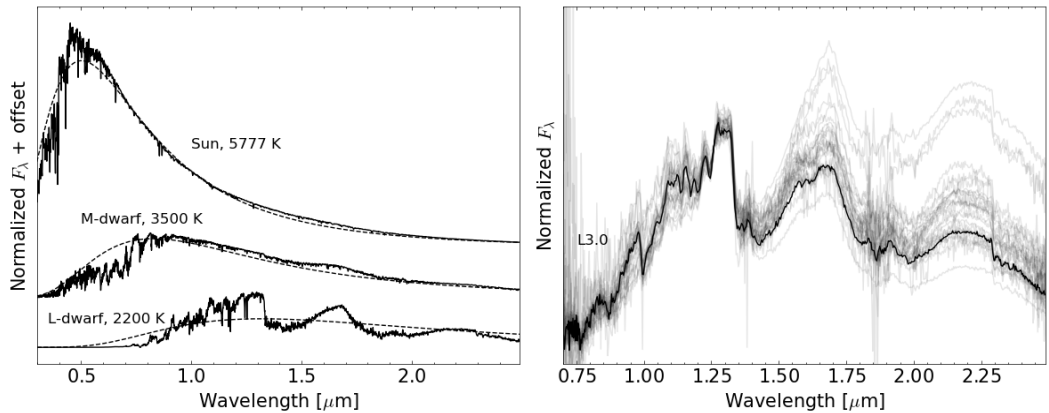

**Fig. 4.** Left: stellar model spectra and black body emission at the same temperature for the Sun, a M-dwarf, and a L-dwarf. Right: Spectra of various L-dwarfs, all classified as L3. The spectra come from the SpeX Prism Spectral Libraries, maintained by Adam Burgasser (Burgasser & Splat Development Team 2017). The black spectrum is the standard 2MASSW J1506544+132106 (Burgasser 2007). Courtesy Thomas Ravinet.

Low mass stars host exoplanets, and are ideal targets for searches for potentially habitable terrestrial planets: e.g. TRAPPIST-1 (Gillon et al. 2016, 2017), Proxima Cen (Anglada-Escudé et al. 2016), Ross 128 (Bonfils et al. 2018). The lowest-mass brown dwarfs more closely resemble the gas giant planets than stars (see Figure 1 by Marley & Leggett 2009) and therefore serve as laboratories for understanding gas-giant extrasolar planets as well as the faint end of the star formation process. When these low-mass brown dwarfs are close enough and bright enough to be observed spectroscopically, their atmospheres are much easier to study than the ones of similar exoplanets, hampered by the light of the host star.

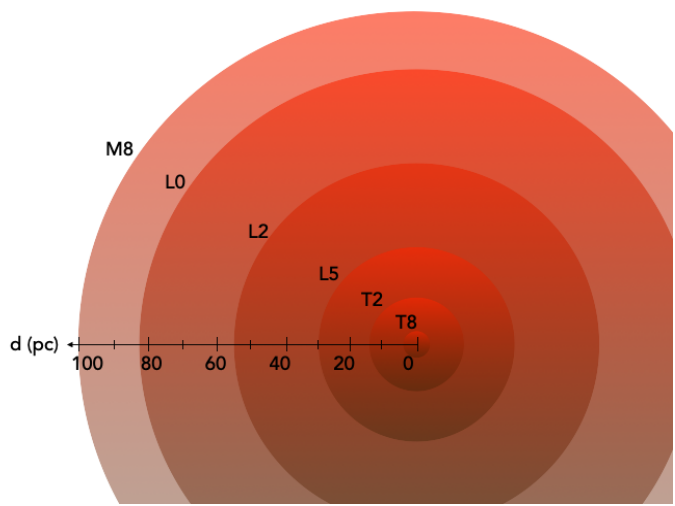

**Fig. 5.** Distance limits vs spectral type assuming the Gaia magnitude limit G=20.7 estimated from Smart et al. (2017).

Hence their study has relevant implications for stellar, exoplanetary, and Galactic astronomy.

## 4  Ultra-cool dwarfs in Gaia

*Gaia* provides the means to uncover ultra-cool dwarfs through astrometric, rather than purely photometric, selection. They can be selected from their locus in the color absolute magnitude diagram. *Gaia* holds the promise of a truly volume-complete sample, with distance limit depending on the spectral type (Figure 5). Sarro et al. (2013) estimated the expected end-of-mission number of UCDs in the Gaia archive: more than 40 000 objects, 600 objects between L0 and L5, 30 objects between L5 and T0, and 10 objects between T0 and T8.

Later on, (Smart et al. 2017) compiled the Gaia ultra-cool dwarf sample (GUCDS), a catalogue of known L and T dwarfs spectroscopically confirmed at that time, containing 1010 L and 58 T having a *Gaia* $G$ predicted magnitude smaller than 21.5. They cross-matched these known UCDs with *Gaia* first data release (DR1 Gaia Collaboration et al. 2016) and identified 319 L and 2 T. Although incomplete, this first sample was used by the Data Processing and Analysis Consortium (DPAC) as a starting point to develop pipeline for parameter estimation purposes based on the low resolution *Gaia* RP spectra (see Carrasco et al. 2021; De Angeli et al. 2023). Gaia Collaboration et al. (2018a) found 443 L and 7 T among selected precise data from the second release (DR2, Gaia Collaboration et al. 2018b). With a slightly different selection, Gaia Collaboration et al. (2018b) identified 647 L and 16 T, as well as 3671 M7 to M9 dwarfs. Finally 4767 known M7 to M9, 1061 L and 16 T are identified in the third data release (DR3, Gaia Collaboration et al. 2023b, Richard Smart, private communication).

This sample constitutes an unprecedented sample, with precise distance determination, to define absolute magnitude vs color, and vs spectral type relations (see Table 1 from Reylé 2018). It provides strong constraints to evolutionary models (Figure 6, left panel). Finally, its locus in the precise $M_G$ vs $G - G_{\rm RP}$ diagram offers an obvious way to identify new UCD candidates (Figure 6, right panel).

To select robust UCD candidates, Reylé (2018) used *Gaia* DR2 data filtered following Gaia Collaboration et al. (2018a) who selected the most precise data (in parallax and photometry, but also handling the extinction rigorously), without trying to reach completeness. The selection in done in the $M_G$ vs $G - G_{\rm RP}$ diagram based on the locus of the known UCD sample. As explained by Evans et al. (2018), faint objects may have spurious $G - G_{\rm RP}$ colors because the measured RP (as well as BP) flux may include a contribution of flux from bright sky background (e.g., in crowded regions or from a nearby source). Evans et al. (2018) defined a filter based on the consistency between the flux in the $G$, $G_{\rm BP}$, and $G_{\rm RP}$ bandpass to remove objects with spurious colors. However the use of $G_{\rm BP}$ is not appropriate for UCDs because these faint and red objects have a low flux in this band. Reylé (2018) defined a new empirical filter based on the 2MASS J band, excluding objects with spurious colors, but retaining the low-mass objects (in particular L-dwarfs). This selection let to numerous new candidates, 14 176 M7 to M9 and 488 L (all earlier than L5). They are evenly distributed across the sky, filling in missing populations in the Southern hemisphere and Galactic plane (Figure 7). Their distance shows that the local census is still incomplete. At 25 pc, the number of new candidates equals the number of known objects retrieved in *Gaia*DR2. Beyond 30 pc, the number of new candidates exceeds that of the known sample. Using the same technique, 2879 additional candidates were identified in the GCNS based on *Gaia*EDR3.

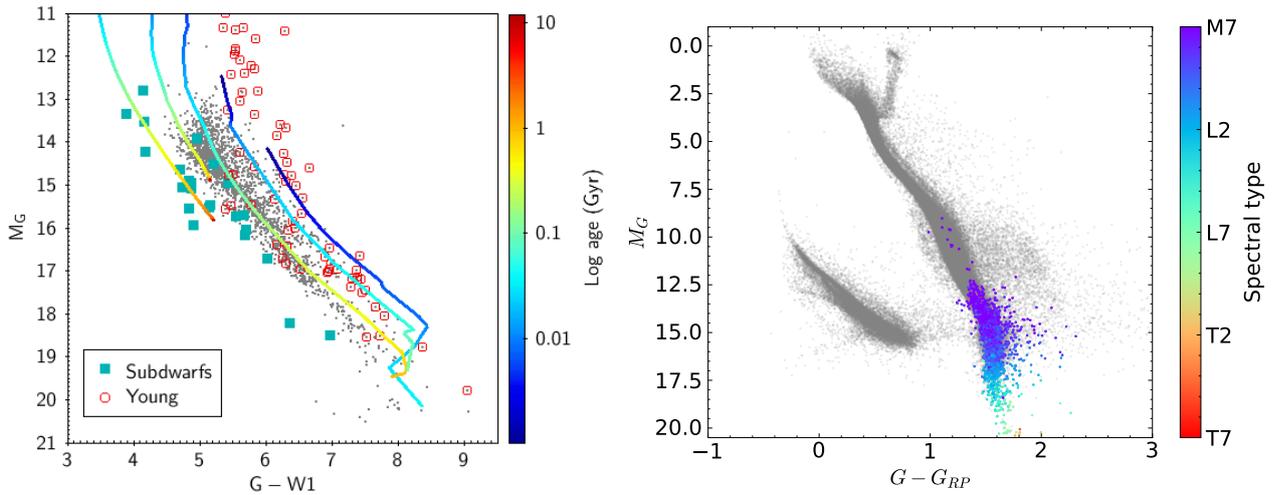

**Fig. 6.** Left: Known UCDs identified in *Gaia*DR2. The symbols depict the objects listed as potential subdwarfs or young candidates. The lines are evolution tracks at different masses and metallicities (from right to left : 0.1, 0.2, 0.5, 0.9 $M_\odot$ at solar metallicity, and 0.83 $M_\odot$ at metallicity -1). The circles on the tracks are colored by the age in Gyr. The evolutionary models are computed by Baraffe et al. (2015) that consistently couple interior structure calculations with the BT-Settl atmosphere models (Allard et al. 2013). Right: locus of known UCDs in the $M_G$ vs $G - G_{RP}$ diagram, colored by spectral type.

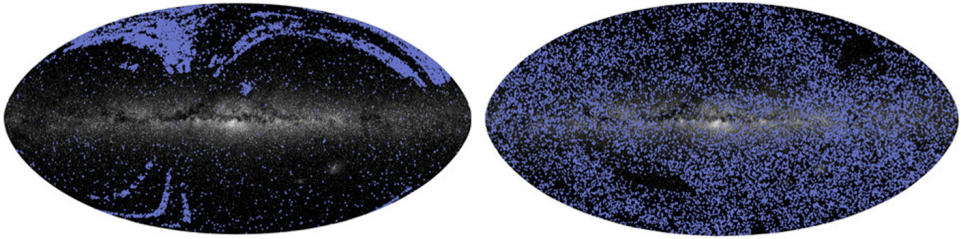

**Fig. 7.** Sky distribution of known UCDs (left) and new candidates found in *Gaia*DR2 (right).

The *Gaia* third data release (DR3, Gaia Collaboration et al. 2023b) offered in addition the opportunity to use low-resolution spectra to refine and widen the selection. The ESP-UCD (*Extended Stellar Parametrizer Ultra Cool Dwarfs*) module infers the effective temperature from the shape of the RP spectrum (Figure 8). 94 158 UCD candidates with $T_{\text{eff}}$ estimates below 2700 K were identified by Sarro et al. (2023). The readers are also referred to the lecture noted by Creevey (2024) who presents the data products from *Gaia*DR3 that can be exploited for stellar physics.

## 5 Characterization

### 5.1 Astrophysical parameters

Several efforts have been made and are still on-going for spectroscopic follow-up of new UCD candidates in order to confirm their nature and further characterize them. In general the spectroscopic follow-up requires the use of 4-meter facilities.

Near-infrared spectroscopic follow-up of 60 nearby (closer than 50 pc) UCDs candidates with the SOFI intrument at the New Technology Telescope (NTT) in la Silla observatory, are presented by Ravinet et al. (2024). Their spectral type, derived from template-matching (Figure 9, left panel) using the SpeX Prism

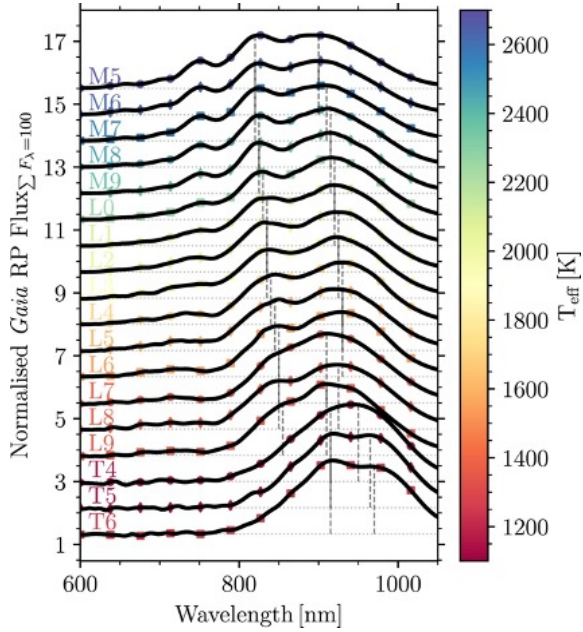

**Fig. 8.** Normalized median RP fluxes for M5 to T6 spectral type, indicated by labels and coloured by effective temperature. Vertical dashed lines indicate the position of the two primary spectral peaks. All spectra are normalized and linearly offset. From Cooper et al. (2024)

Library Analysis Toolkit (SPLAT, Burgasser & Splat Development Team 2017) is very close (standard deviation of one subtype) to the one expected using their absolute magnitude $M_G$, showing that $M_G$ can be used with confidence as a proxy to select UCDs in *Gaia* data. Ravinet et al (in prep) also made a comparison with synthetic spectra (Figure 9, right panel) computed from various atmospheric models: DRIFT (Witte et al. 2011), BT-Settl-CIFIST and BT-Settl-AGSS (Allard 2014) (assuming different solar abundances Caffau et al. (2011) and Asplund et al. (2009), respectively), and ATMO (Phillips et al. 2020). The retrieved stellar parameters ($T_{\rm eff}$, $\log g$, [Fe/H]) can be quite different depending on the models used: comparing properties of stars obtained with different models must be handled with care. In general, ATMO synthetic spectra give the best fit, although the H-band (1.5 - 1.8 $\mu$m) is often poorly reproduced by all models, in particular at the M/L transition where dust starts to condensate in the atmosphere. It illustrates the difficulty to properly model the effect of dust in these cool atmospheres.

Spectroscopic follow-up permits also to identify binaries from "flux-reversal": binaries whose component straddle the L/T transition show a brighter secondary in the J-band than the H and K-bands (e.g. Cruz et al. 2004; Looper et al. 2008). Such unresolved binaries, named spectral binaries, can be identified by an index-identification technique (Burgasser et al. 2010; Bardalez Gagliuffi et al. 2014), using the spectral deviations between the binaries and templates, such as a stronger methane absorption in the H-band relative to the K-band, the K-band flux peak slightly shifted toward the blue in the binaries, etc (Figure 10).

A new advent came with *Gaia* DR3 that provides astrophysical parameters (Creevey et al. 2023). The "golden sample", which is a homogeneous sample of stars with high-quality astrophysical parameters compiled by Gaia Collaboration et al. (2023a) contains 21 068 UCDs. Figure 11 illustrates in two ways the quality and potential scientific outcome of this sample. The top panel shows the $T_{\rm eff}$ determined from the ESP-UCD module vs the radius determined from the FLAME module. It shows the expected decrease in radius as the temperature decreases with a minimum value at 2000-2200 K, in agreement with the findings of Dieterich et al. (2014), followed by an increase until 1400 K where the trend reverses. Another application is to constrain the characteristics of faint, unseen UCDs that are beyond the *Gaia* magnitude limit but are in binary systems with brighter objects. 11 multiple systems were identified from their common distance and proper motion (after allowing for orbital motion). Assuming that the UCD has the same age as the primary, the absolute magnitude vs age diagram, superimposed with evolutionary tracks (Figure 11, bottom rpanel) gives some indication on the stellar or substellar nature of the UCDs.

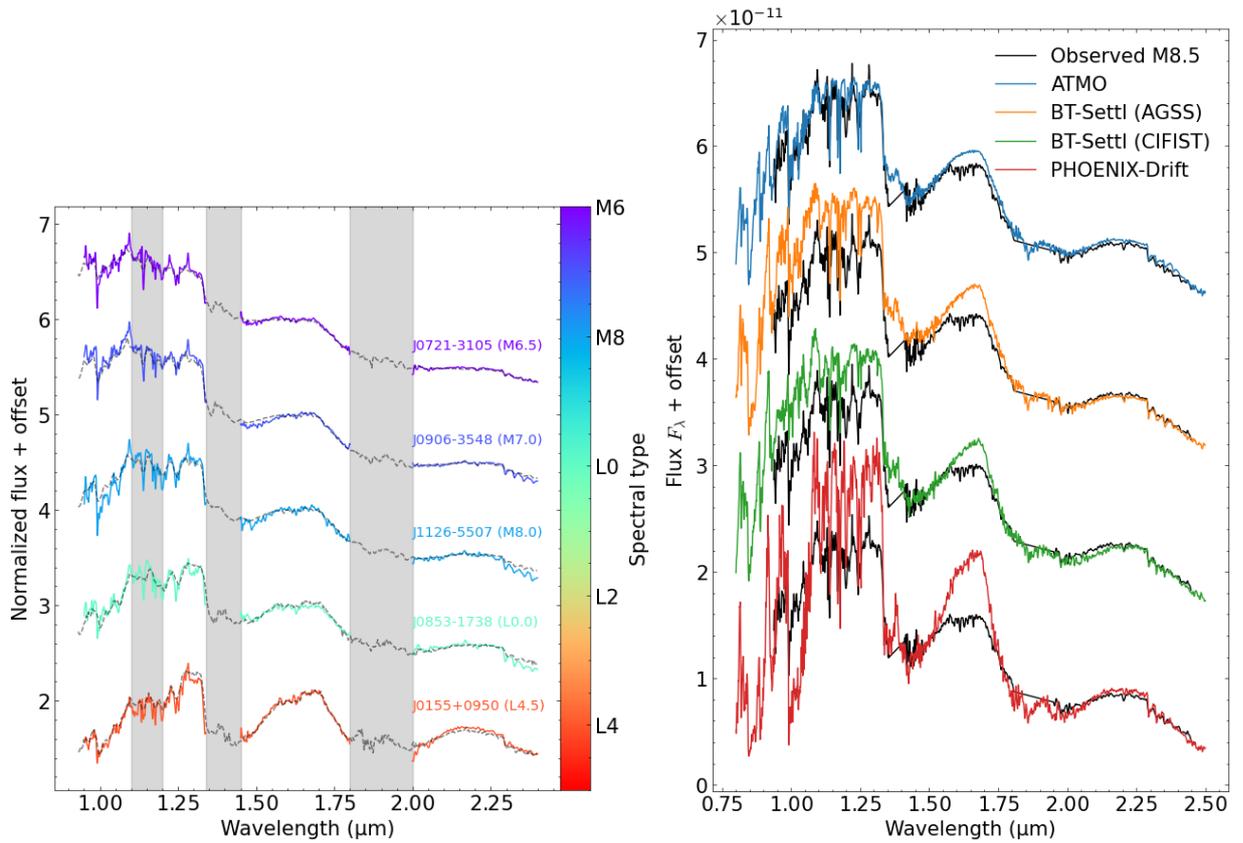

**Fig. 9.** Left: SOFI spectra (filled lines, colored by spectral type) of UCDs, together with standard spectra of the same spectral type. Right: SOFI spectrum of a M8.5 (black) superimposed with best fit synthetic spectra computed from different atmosphere models. Courtesy Thomas Ravinet.

## 5.2 Subdwarfs

Subdwarfs are low-metallicity objects and are typically found to have thick disc or halo kinematics. The name was coined by Gerard Kuiper in 1939, to refer to a series of stars with anomalous spectra that lie between the Main Sequence and the white dwarf sequence in the Hertzsprung-Russel diagram and were previously labeled as "intermediate white dwarfs" (Figure 12, left panel). Solar-metallicity objects are classified as dwarfs, while the more metal-poor stars are classified as subdwarfs (sd), extreme subdwarfs (esd), and ultra subdwarfs (usd), in order of decreasing metallicity (Gizis 1997; Lépine et al. 2007). Because of a decreasing metallicity, TiO opacity decreases in M-subdwarf atmospheres. Less blanketing from TiO bands means more continuum flux radiated from hotter and deeper layers of the atmosphere. The M-subdwarf spectrum is closer to that of a blackbody, and subdwarfs appear bluer (Figure 12, right panel). In L-subdwarfs, the enhanced collision-induced $H_2$ absorptions suppress the K and K-bands.

Because low-mass subdwarfs have lifetimes far in excess of the age of the Galaxy, they are important tracers of Galactic chemical history and representative of the first generations of star formation. To date hundreds of late-type M-subdwarfs (e.g. Lépine et al. 2003; Burgasser et al. 2007; Lépine & Scholz 2008; Lodieu et al. 2012, 2017; Kirkpatrick et al. 2016), and ∼70 L-subdwarfs (e.g. Burgasser et al. 2003; Kirkpatrick et al. 2014; Zhang et al. 2017, 2018) have been discovered with modern sky surveys, often detected from their high proper motion.

Hejazi et al. (2018) used the accurate *Gaia*DR2 parallaxes of ∼1600 high proper motion M-dwarfs and subdwarfs with spectroscopic metallicity measurements to draw the $M_G$ vs $G_{BP} - G_{RP}$ diagram. Stars with different metallicity ranges fall into clearly distinct loci which can be used to develop photometric metallicity calibrations, in particular for metal-poor M subdwarfs.

Zhang et al. (2018) selected L subdwarfs from their locus in optical and near infrared color-color diagrams, and confirmed their nature with spectroscopic follow-up. They computed the tangential velocity for 20 of them

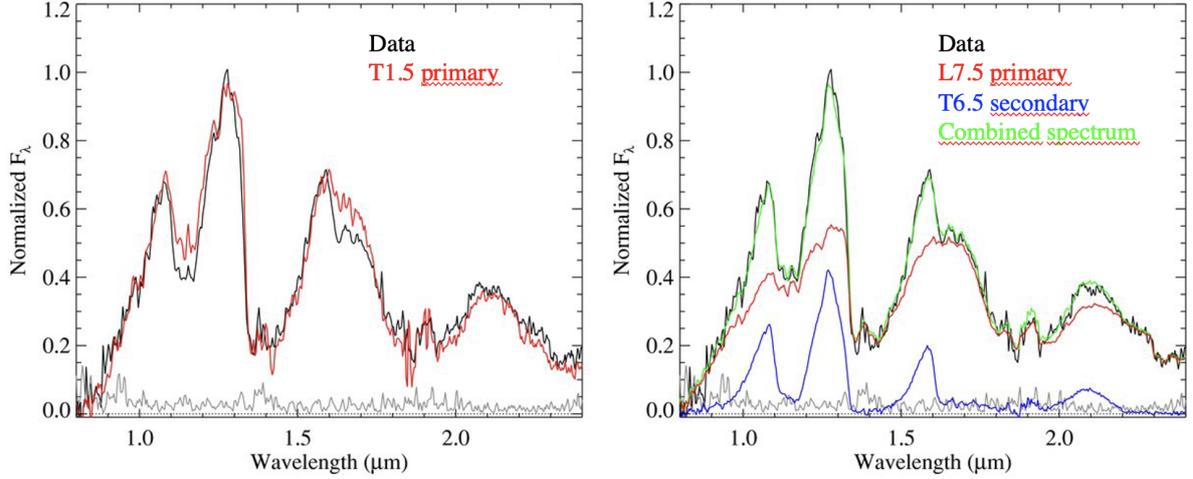

**Fig. 10.** Spectral binary candidate. The left panel shows the best-fit single template (red line) compared to the source spectrum (black line). The right panel shows the best-fit composite (green line), primary (red line), and secondary spectra (blue line). Adapted from Burgasser et al. (2010).

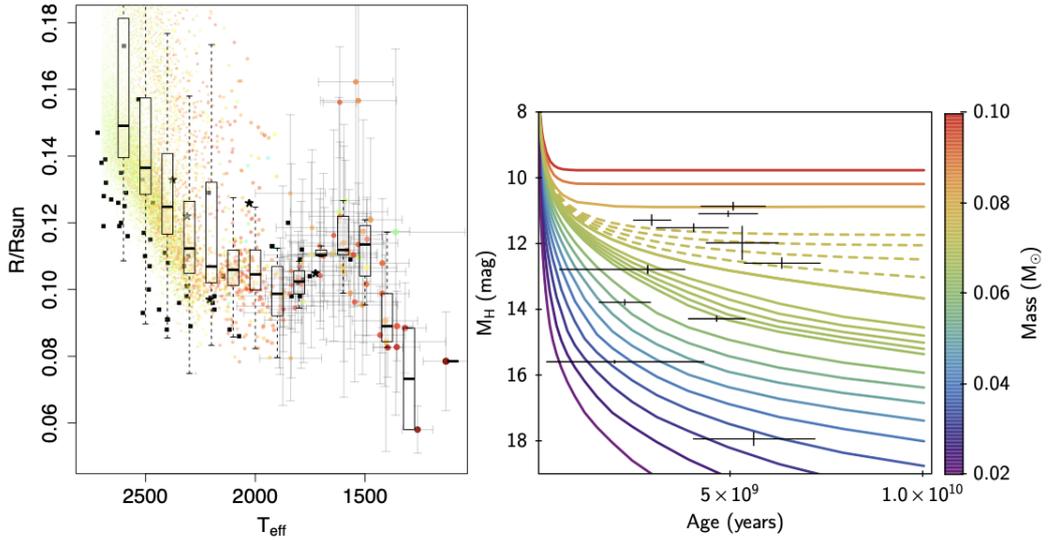

**Fig. 11.** Left: Radii of candidate UCDs in the Gaia golden sample (colored symbols), superimposed to the data of Dieterich et al. (2014) (black symbols). The box plots are calculated within bins of 100 K. Right: Evolutionary tracks and UCD (in binary systems) locations in the H-band absolute magnitude vs age diagram, adopting the companion age. The tracks are color coded by mass. The dashed lines indicate the stellar to substellar transition zone (from 0.072 to 0.075 $M_\odot$). From Gaia Collaboration et al. (2023a)

found in the *Gaia*DR2 catalogue. They found that the sdL subclass mostly have thick disc kinematics, whereas the esdL and usdL subclasses generally have halo kinematics, which is consistent to the esdM/usdM subclasses.

### 5.3 Wide binaries

Multiplicity with *Gaia* can be determined from distance, angular separation and proper motion measurements (e.g. Hartman & Lépine 2020; El-Badry et al. 2021; Gaia Collaboration et al. 2021b; Sarro et al. 2023). Several thousands of UCDs in binary systems have been discovered this way. They are useful benchmarks for testing stellar evolutionary models, and can be used as calibrators for age and metallicity relations from their spectroscopic characterization (e.g. Marocco et al. 2020; Low et al. 2021).

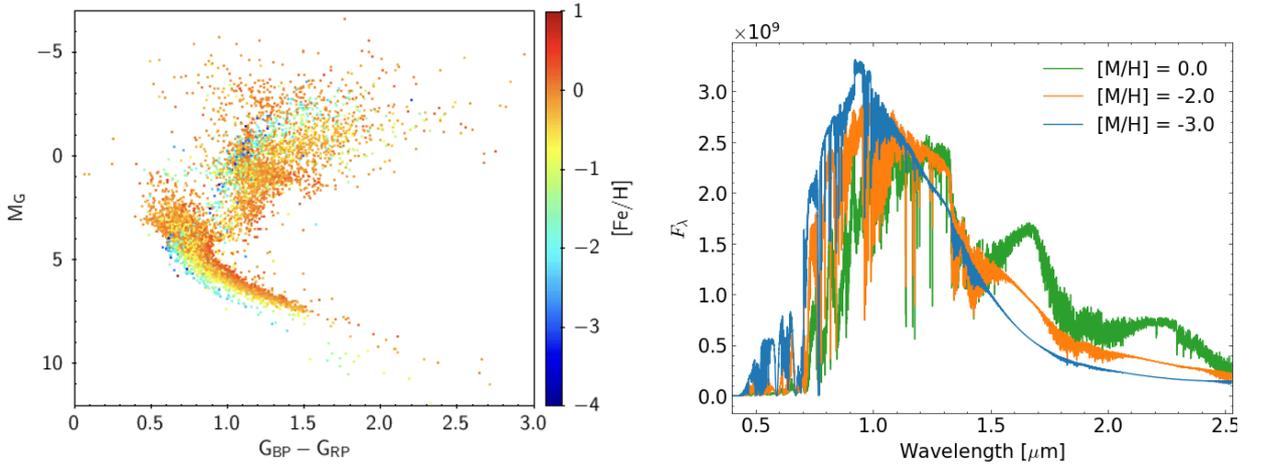

**Fig. 12.** Left: Color absolute magnitude diagram of *Gaia*DR2 stars found in the PASTEL catalogue (Soubiran et al. 2016) wich provides accurate metallicity measurements. The low-metallicity main sequence stars (in blue) form a shifted, lower, sequence. Right: Synthetic spectra computed from BT-Settl atmosphere models with $T_{\rm eff}$ = 2600 K, $\log g$ = 5, and various metallicity values.

Zhang (2019) discovered the first wide M1 + L0 extreme subdwarf binary, Gaia J0452-36AB. Its kinematics is compatible with the halo population, its metallicity is about -1.4, and the projected separation is 15 282 AU. How such old wide systems can survive is still an open question, as well as other ultra-wide systems (González-Payo et al. 2023).

### 5.4  Young associations

The selection of members of young clusters or associations has become trivial with *Gaia*, but also the discovery of new groups, using the parallax, position, and motion of stars (e.g. Cantat-Gaudin et al. 2019; Galli et al. 2019, 2020a,b; Mužić et al. 2022; Tarricq et al. 2021, 2022; Sarro et al. 2023). *Gaia* is sensitive to 30 $M_{\rm Jup}$ up to about 500 pc in young groups. To summarize the conclusions of these numerous studies of young clusters in the solar neighborhood, the mass function in the substellar regime does not vary much, at least within the error bars (Hervé Bouy, priv. comm.).

*Gaia* has furthermore enabled to measure dynamic ages for some of these regions, which is very useful because age is often poorly constrained in these young regions, since evolutionary models are not very reliable for such early ages. This methodology uses the present 3D positions and 3D velocities of individual stars and computes the stellar orbits back in time with a Galactic potential. The dynamical traceback age is the time when a group of stars was most concentrated in the past, that is, when the size of the group was at its minimum (Figure 13). These model-independent ages are therefore particularly interesting. This is particularly important in the context of brown dwarfs whose brightness is very dependent on age and varies rapidly at the beginning of their life, leading to large uncertainties. Miret-Roig et al. (2022) showed that Upper Scorpius and Ophiuchus groups share a common origin. The proposed star formation scenario is likely a result of stellar feedback from massive stars, supernova explosions, and dynamic interactions between stellar groups and the molecular gas.

Using the sample of 94 000 UCD candidates (selected from their *Gaia* $T_{\rm eff}$), Sarro et al. (2023) performed group identification using a hierarchical mode association clustering classification algorithm. Detection and characterization of overdensities are done in the space of celestial coordinates, projected velocities, and parallaxes (Figure 14). They compared the medium RP spectrum in $T_{\rm eff}$ bins for sources outside these associations (Main Sequence) and in several clusters. There is a systematic trend of increasing band depths as the association becomes older. By 10 Myr, the RP spectra of young and Main Sequence sources becomes undistinguishable. This statement can be used as a basis for the indication of youth in the forthcoming *Gaia* spectra.

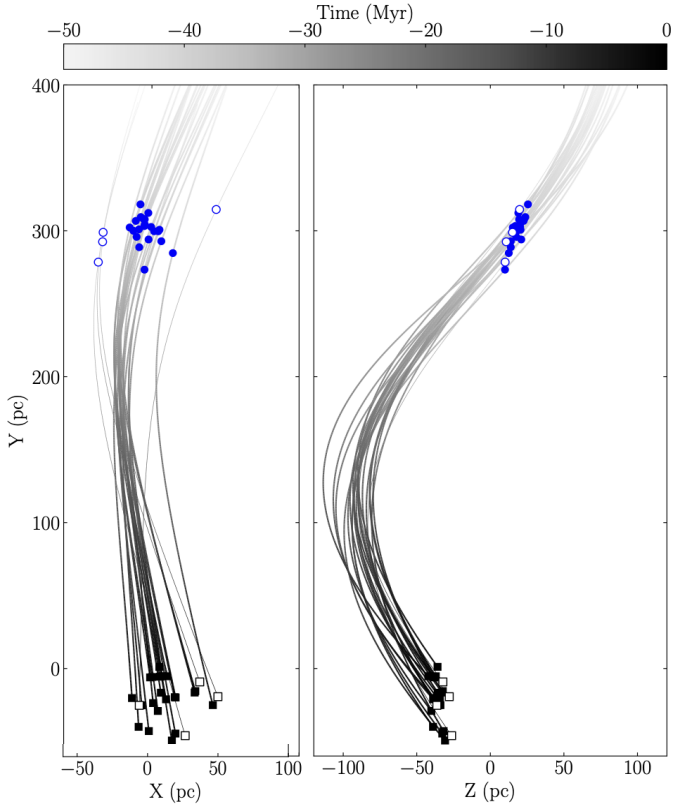

**Fig. 13.** 2D projection of the stellar orbits from 25 stars in the Tucana-Horologium young stellar association. The black squares indicate the present-day position of the stars and the blue circles denote the stellar positions at birth time, 38.5 Myr ago. The orbits are color-coded based on the traceback time. Adapted from Galli et al. (2023).

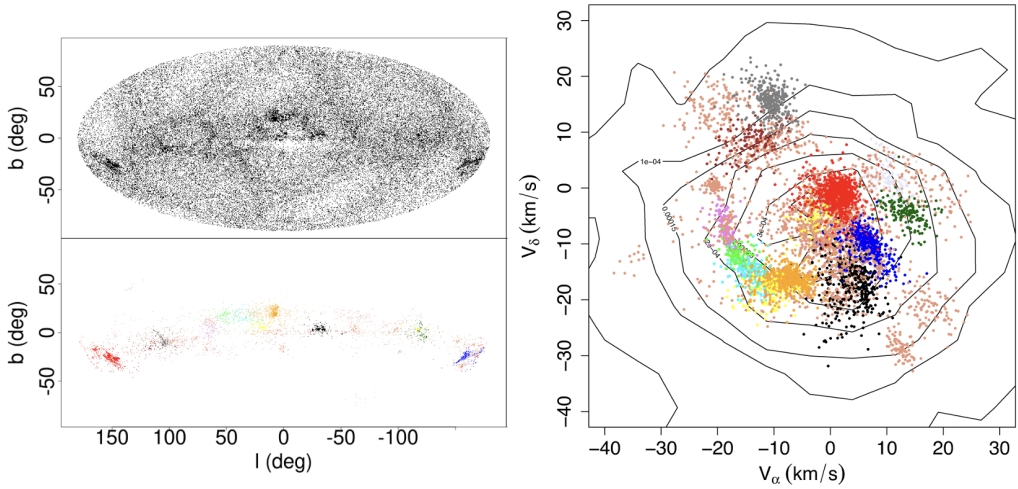

**Fig. 14.** Left: Distribution in Galactic coordinates of the *Gaia*UCD catalogue using the Aitoff projection. Clusters with more than 10 members are depicted with colors. Right: Distribution in the space of tangential velocities of the sources in clusters with more than 10 members with the same color code. From Sarro et al. (2023).

## 6 Towards a complete local census

Because *Gaia* revealed a huge number of UCDs, it constitutes a step towards a complete local census. The nearby sample is particularly important for the ultra-cool dwarfs which are the lowest-mass, coldest, and faintest products of star formation, making them difficult to study at large distances. Having a volume-complete sample, with good statistics, is crucial to compute precise bias-free densities, and to therefore determine luminosity, and mass functions that will bring strong constraints on stellar and substellar formation theories.

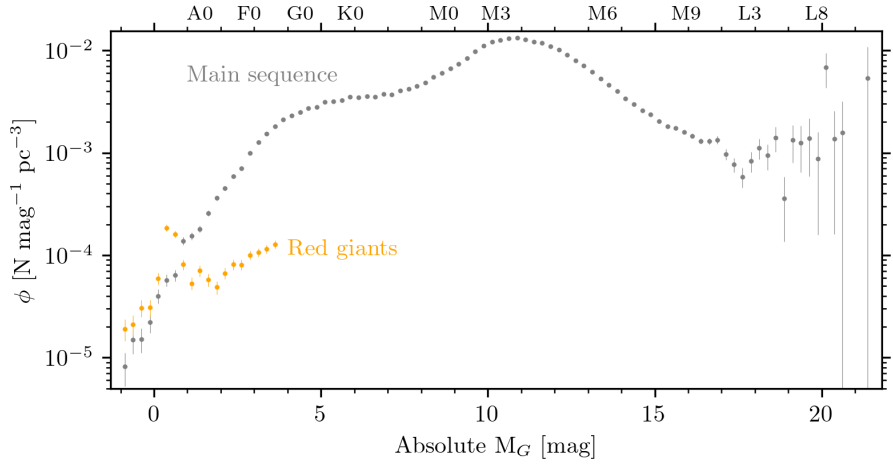

**Fig. 15.** Luminosity function of the GCNS, with a 0.25 bin, in log scale. The upper full curve plotted in grey shows main sequence stars. The small lower partially orange curve shows giants stars.

For substellar objects with no obvious mass-to-luminosity relation, the luminosity function is simulated assuming different initial mass fonctions and birth rates. Thus the comparison with the observed luminosity function allows to disentangle between the different formation scenarios, only if error bars are small enough.

Field stars and brown dwarfs are several Gyr in average. Brown dwarfs depopulate rapidly earlier spectral types to go to later ones. As a consequence, a minimum is expected in the density at the stellar substellar boundary, as predicted by simulations (Burgasser 2004; Allen et al. 2005). Observations may indicate such a minimum in the luminosity function (Cruz et al. 2007; Bardalez Gagliuffi et al. 2019). The GCNS is an exquisite dataset from which to derive the local luminosity function. This is possible for the first time using volume-limited samples with parallaxes not derived from photometric measurements that are affected by related biases (Eddington or Malmquist), from bright stars down to the substellar regime. A clear dip around the L3 spectral type is visible, that is interpreted as the stellar-substellar boundary.

## 7 Conclusions

There are numerous ultra-cool dwarves revealed in the successive *Gaia* data releases. They offer high number, high precision, information on these objects. Photometry and astrometry are so precise that the locus in the color absolute magnitude diagram can give indication on the nature of the object: youth, binary, low metallicity. In addition, *Gaia* offers kinematical and astrophysical parameters parameters for a large part of the catalogue. A well-characterized sample with spectroscopic follow-up is powerful to test (sub)stellar models (evolution, interior, atmosphere). A well-characterized and complete volume-limited sample provides luminosity and mass functions free of biases that plagued previous determinations, provides new insights on the stellar-substellar limit, provides strong constraints on stellar and substellar formation theories. We expect the forthcoming *Gaia* data releases and their additional parameters to continue the story of these puzzling and elusive objects.

This work has made use of data from the European Space Agency (ESA) mission *Gaia* (https://www.cosmos.esa.int/gaia), processed by the *Gaia* Data Processing and Analysis Consortium (DPAC, https://www.cosmos.esa.int/web/gaia/dpac/consortium). Funding for the DPAC has been provided by national institutions, in particular, the institutions participating in the *Gaia* Multilateral Agreement.